# Prépublication

Peter Stockinger

# De la donnée au corpus : enjeux sémiotiques et documentaires des archives audiovisuelles.

**Préface** à l'ouvrage collectif :

----------------------------------------------
Institut national des langues et civilisations orientales (INALCO) Équipe « Pluralité des langues et des identités » (PLIDAM)

Lambach - Paris 2022

## Table des matieres

# Introduction

Cet ouvrage réunit dix contributions au colloque *Le corpus audiovisuel – Quelles approches ? Quels usages*. Ce colloque s'est tenu en juin 2019 à l'Inalco (Institut National des Langues et Civilisations Orientales) à Paris et a été organisé par Louise Ouvrard et Peter Stockinger de l'équipe Plidam (EA 4514 – Pluralité des langues et des identités : didactique, acquisition et médiations). Depuis plusieurs années déjà, l'équipe Plidam est activement engagée dans une série de recherches portant sur la constitution et l'exploitation de corpus audiovisuels, visuels et sonores, documentant d'une part la recherche en didactique des langues et des cultures, et d'autre part les enquêtes menées en communication interculturelle sur les pratiques et activités quotidiennes.

Dédiées aux corpus audiovisuels, plus particulièrement à leurs utilisations et à leurs exploitations dans le cadre de projets de recherche concrets, les contributions réunies dans cet ouvrage se positionnent dans ce (relatif) nouveau domaine foisonnant d'idées, de propositions théoriques, méthodologiques et techniques que sont les humanités numériques. Mise à part la spécificité des objets et des questions qui sont abordées par chacune des contributions, il nous semble qu'elles abordent, d'une manière plus ou moins directe ou indirecte, cinq thématiques transversales. Les voici :

1. le statut et la structure d'une donnée textuelle lato sensu (incluant, bien sûr, les données visuelles, audiovisuelles et sonores) ;
2. la valeur documentaire d'une donnée textuelle/d'un corpus de données textuelles ;
3. la distinction entre fonds de données, corpus de données et archives de données ;
4. la « sémantisation » des données textuelles ;
5. l'instrumentation du travail avec un corpus de données, i.e. l'usage d'environnements et de logiciels appropriés.

Considérons rapidement ces cinq thématiques qui nous semblent posséder en effet une importance tout à fait centrale pour les recherches en humanités numériques qui s'intéressent aux *archives numériques*.

# 1) Statut et structure d'une donnee textuelle lato sensu

Dans un certain sens peut être considérée comme une donnée textuelle toute trace qui renseigne ou qui semble apporter des renseignements sur un domaine de connaissance. Ainsi, la donnée textuelle peut se matérialiser dans tout support média imaginable : documents écrits et imprimés (bien sûr) mais aussi témoignages oraux, photographies, dessins, enregistrements sonores, captations vidéo, objets et scènes 3D, etc.

Ce sont des données textuelles dans la mesure où elles expriment, mettent en scène – comme le texte écrit – une certaine vision, une conception, i.e. un sens particulier du domaine qu'elles documentent. Cette mise en scène d'une certaine vision du domaine n'est pas limitée, contrairement au texte écrit, à un seul système de signes d'expression (i.e. à l'écrit) ni aux formats classiques de documents écrits ou imprimés (par exemple : à l'article, au journal, au livre, …). Au contraire, elle s'appuie d'une part sur toutes les *modalités d'expression* accessibles à la perception humaine et d'autre part sur tous les *formats* (image, film, document sonore, …) organisant l'expression d'un contenu, d'un sens par d'autres moyens que l'écrit. Toute cette diversité de données constitue, en effet, des *textes* dans la mesure où elles fournissent des informations sur un domaine, en offrent un accès à sa compréhension, à son appropriation, à son utilisation et, enfin, à sa critique.

La conséquence directe de ce constat est que cette diversité doit être traitée par et dans un même cadre théorique et méthodologique, à l'aide des mêmes méthodes. On comprend ainsi mieux l'importance d'une *sémiotique* comprise comme une approche qui est supposée offrir ce cadre théorique et méthodologique de nature transdisciplinaire.

## 2) La valeur documentaire d'une donnee textuelle

En gros, la *valeur documentaire* d'une donnée textuelle lato sensu renvoie à la *pertinence* que possède une donnée textuelle en particulier et un corpus de données textuelles en général pour documenter d'une manière adéquate un domaine de connaissance.

Du point de vue de l'*utilisateur* d'un corpus de données textuelles déjà existant, il s'agit de savoir comment *évaluer* et *reconnaître* ladite valeur documentaire d'une don- née textuelle particulière qui fait partie du corpus servant à appréhender – à analyser et à comprendre – un domaine de connaissance. Du point de vue du *producteur* d'un corpus, la question se pose plutôt en termes de choix argumenté, de sélection raison- née des « bonnes » données textuelles, de données textuelles qui sont suffisamment pertinentes pour figurer dans un corpus documentant un domaine donné.

Ces deux interrogations renvoient d'abord au problème épistémologique plus fonda- mental de la *perspectivité épistémique inhérente*, de l'*indexicalité* de toute connaissance (y incluse de la connaissance scientifique ou à prétention scientifique). Cependant, contrairement à une approche précritique, naïve ou dogmatique, une approche scientifique ou critique est supposée thématiser son *point de vue* et argumenter le *choix des méthodes* servant à la constitution d'un corpus qui documente un domaine de connaissance. Ce n'est qu'en connaissance de ce préalable critique que la valeur documentaire d'une donnée textuelle en particulier, et d'un corpus de données en général, peut être appréciée. Plus concrètement, on rencontre ici toutes les interrogations sur la provenance et le statut d'une donnée textuelle (ou d'un corpus de données textuelles), sur l'historique et les modalités de la production d'un corpus de données, sur la « biographie » (la génétique) d'une donnée textuelle, sur sa substituabilité ou encore sur l'existence de variantes qui dérivent d'une donnée textuelle originale.

Cela dit, la valeur documentaire d'une donnée textuelle n'est pas seulement conditionnée par la perspectivité épistémique inhérente à toute connaissance ; elle est aussi tributaire de la condition *pragmatique* de sa place dans un projet concret. La notion de projet comprend un ensemble d'activités qui, en fonction d'un *objectif* particulier, s'intéresse à un domaine particulier, le domaine étant documenté par un ensemble de données textuelles lato sensu appelé *corpus*.

Un domaine particulier, comme par exemple une époque, une région ou une population, peut intéresser une grande diversité de projets : projets de recherche scientifique, projet de communication et de valorisation, projet d'enseignement et de formation, projet de création d'un patrimoine et d'une mémoire, … Mais, bien entendu, selon le type ou le genre de projet, le *regard* sur le domaine, l'intérêt qu'on lui apporte, la compréhension qu'on peut en avoir, diffèrent d'une manière plus ou moins radicale. Un projet de présentation d'une région aura recours à des jeux de données qui varient d'une manière plus ou moins significative selon son objectif précis : pédagogique dans le cadre d'un projet d'enseignement, social dans le cadre d'un projet de politique local, financier et économique dans le cadre d'un projet d'investissement, publicitaire dans le cadre d'un projet touristique.

Une conséquence directe est celle de la *documentation appropriée* d'un domaine en fonction du type ou du genre de projet. Autrement dit, se pose ici le problème de la *valeur* d'une donnée textuelle en particulier et d'un corpus de données textuelles en général en fonction du projet dans le cadre duquel il est constitué et utilisé. Se pose également le problème de la *variabilité* d'un corpus, de sa représentativité pour et de sa ré-employabilité dans différents projets. On rencontre entre autres les questions suivantes : quels sous-ensembles de données peuvent être utilisés dans différents projets ? ; quelles données doivent être considérées comme strictement limitées à un projet ? ; quelles autres données possèdent une employabilité plus large ? ; comment préciser la représentativité empirique d'un corpus en fonction d'un projet et, par conséquent, quelle est la volumétrie à respecter ? ; enfin, peut-on se contenter d'un corpus fermé et figé ou faut-il s'assurer de son enrichissement, de son renouvellement et si oui, par quels types de données et en l'assurant comment, par quels moyens ?

## 3) LA DISTINCTION ENTRE FONDS DE DONNEES, CORPUS DE DONNEES ET ARCHIVES DE DONNEES

Il convient de distinguer d'abord entre un *fonds de données textuelles* à la disposition d'un acteur (par exemple d'un chercheur ou d'une communauté de chercheurs) pour appréhender un domaine d'expertise ou de connaissance et le *corpus de données textuelles* qui se présente comme une sélection de données textuelles provenant du fonds et dont la sélection se fait en fonction de la nature et de l'objectif d'un projet.

En d'autres termes, un fonds de données peut nourrir toute une diversité de corpus mais n'est pas (obligatoirement) utilisable, tel quel, par un projet particulier. Ainsi, l'archive ouverte HAL[1] offre aux innombrables projets scientifiques un fonds inestimable de données (sous forme de dépôts réalisés par la communauté des chercheurs). L'ensemble de ce fonds cependant, n'est bien évidemment pas pertinent pour un projet de recherche concret et circonscrit.

Prenons aussi l'exemple de Nakala qui est un service du TGIR (Très Grande Infrastructure de Recherche) Huma-num[2] du CNRS offrant aux chercheurs en sciences humaines et sociales un environnement technique, scientifique et humain pour le dépôt, la publication en ligne et la valorisation de leurs données de recherche. Les données déposées par la communauté des chercheurs peuvent être des corpus entiers, des éléments de corpus ou encore des jeux de données ou des données isolées qui possèdent une valeur plutôt illustrative. Les milliers de données se trouvant archivées dans l'entrepôt Nakala constituent un fonds riche et diversifié qui témoigne d'un grand nombre d'activités et de projets de recherche en sciences humaines et sociales.

En respectant le cadre légal, on peut les utiliser ou, plutôt, les réutiliser, dans le cadre d'un projet concret. Bien entendu, c'est au chercheur non seulement de sélectionner les éléments textuels dont il a besoin pour se constituer un corpus pertinent par rapport à son projet mais aussi de produire les *arguments* montrant le bien-fondé de son choix.

Autrement dit, la sélection de données dans un fonds existant ne peut être réduite à un simple acte technique. Pour que la sélection fasse sens, pour qu'elle soit motivée, elle doit être explicitée, argumentée. Il est tout à fait indispensable de rendre compte de cette opération cognitive. C'est elle, et elle seule, qui nous fournit les informations nécessaires pour distinguer *corpus* intentionnellement élaboré et *fonds de données* existant, pour distinguer corpus de données prétendant fournir une documentation adéquate d'un domaine de connaissance et jeu de données simplement illustratif. Elle nous fournit également les critères dont nous avons besoin pour évaluer l'adéquation d'un corpus en fonction des objectifs du projet dans lequel il est utilisé.

Enfin, considérons encore rapidement la troisième notion introduite ci-dessus, celle d'*archives de données*. Cette notion a connu une popularité considérable depuis deux décennies, aussi bien dans le cadre des recherches en humanités numériques que dans celui traitant de la problématique de la constitution de patrimoines de connaissance. Nous avons argumenté ailleurs (Stockinger 2011, 2012, 2015) le fait que la notion d'« archives numériques » revêt deux problématiques différentes mais indissociablement liées :

1. les archives au sens d'une *banque de données* (ouvertes) ;
2. les archives au sens d'une *ressource ou d'un ensemble de ressources* (d'un centre de ressources).

---

[1] Cf. le site HAL : https://hal.archives-ouvertes.fr/
[2] Pour plus d'informations, *cf.* le site de la TGIR Huma-num : https://www.huma-num.fr/

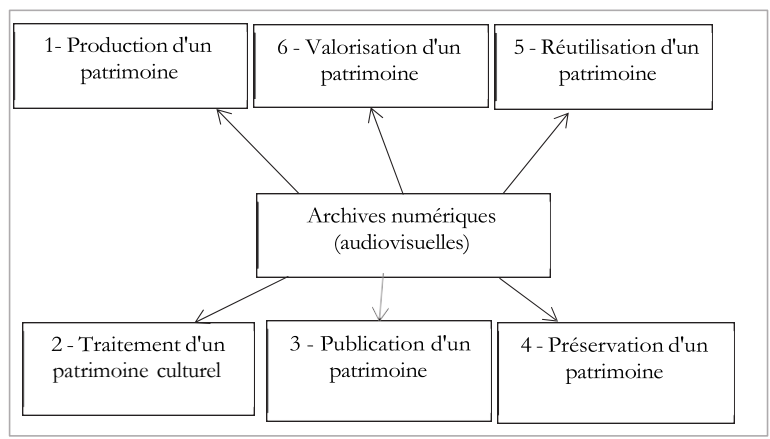

**Fig. 1** : schéma illustratif des principales activités réalisées dans l'écosystème des archives numériques

Les archives au sens d'une *banque de données* (ouvertes, d'accès libre, …) font partie d'un *projet d'archivage* spécifique, se présentent sous forme de collections de données numériques (composant le fonds – fermé ou ouvert – d'une archive) et peuvent être *potentiellement* l'objet de toutes sortes d'exploitations pratiques.

Dans le deuxième sens, les archives numériques servent d'objet de recherche théorique et appliquée en vue, par exemple, d'améliorer les *accès aux données archivées*, d'avancer sur les *terrains de la sémantisation et de l'interconnexion* des données archivées (*cf.* ci-dessus) ou encore de mieux contrôler le processus de *l'éditorialisation* et de la *visualisation* (*lato sensu*) des données et des métadonnées.

Le diagramme ci-après identifie quelques activités majeures qu'offre une plateforme 2.0 ou 3.0 et qui façonnent tout projet d'archivage (*cf.* Stockinger 2010, 2011, 2012). Sans entrer dans plus de détails, ajoutons qu'ensemble ces activités caractéristiques de tout projet d'archivage montrent la pertinence et l'intérêt de la *vision sémiotique* (du « semiotic turn », *cf.* Stockinger, 2015 ; *cf.* aussi Craven 2008 qui adopte un point de vue similaire) pour la recherche sur les archives numériques : d'une manière générale, le *fonds* d'une archive peut être considéré comme un ensemble (quantitativement plus ou moins important) de *données textuelles* au sens large du terme qui à la fois documentent, conservent et transmettent le *discours* d'un *acteur social* (d'une organisation, d'un groupe social, d'une communauté, voire d'une personne) sur un domaine, un objet, un lieu ou une époque[3].

---

[3] Cette vision fait référence aux travaux antérieurs de Michel Foucault sur la fonction des archives dans la constitution d'une *rationalité* (par exemple *médicale* ou *pénale*, en ce qui concerne les recherches de Foucault), *i.e.* sur la constitution d'un ensemble de savoirs et savoir-faire formant un ensemble de standards (de traditions, normes ou règles) auxquels se réfère un corps social et/ou professionnel (médecins, psychiatres, pénalistes…) pour exercer son métier.

# 4) La question de la « sémantisation » d'une donnée/d'un corpus de données textuelles

La problématique que nous désignons par l'expression « sémantisation des données » renvoie directement à la question de la *valeur documentaire* d'une donnée ou d'un corpus de données. Elle s'intéresse plus particulièrement à la *manipulation intentionnelle* (i.e. au *traitement*, si on préfère ce terme plus neutre), à la fois *sémiotique* et *technique*, d'une donnée textuelle ou, plutôt, d'un corpus de données textuelles en vue de le faire correspondre aux objectifs d'un projet donné.

Sous le sigle de la « sémantisation des données » (qu'on peut rapprocher de l'expression « smart datas »), se posent trois questions tout à fait complémentaires :

1. Comment identifier, localiser – dans la masse des données numériques – les données qui possèdent une pertinence, une *valeur* pour un acteur social donné, qui forment, autrement dit, des *signes potentiels* pour un acteur social en quête d'informations ou de connaissances sur un domaine, un objet documenté par ces données ?
2. Comment décrire, analyser la *structure sémiotique* – le *contenu* et l'*expression* – de ces données pour pouvoir expliciter *leur valeur (potentielle)* pour un acteur social en quête d'informations ou de connaissances sur un domaine, un objet particulier ?
3. Comment expliciter, enfin, le *positionnement* des données les unes par rapport aux autres, i.e. comment expliciter les *rapports intertextuels* entre un ensemble de données qui constituent (hypothétiquement) les éléments signifiants – le paysage textuel ou encore le *textscape multimodal* – d'un domaine ou d'un objet au sujet duquel un acteur social est en quête de connaissance ?

Ces trois questions renvoient à la problématique plus générale de la *valorisation* de données ou encore à celle de la création de *chaînes de valeurs* (pas obligatoirement restreintes aux organisations économiques – bien loin de là !) à partir de ou sur la base d'un fonds de données stockées, par exemple, dans des entrepôts ou archives numériques. Technologiquement parlant, il s'agit ici de la conception, implémentation et utilisation de *plateformes d'outils*, de *méthodes* et de *bonnes pratiques* pour :

1. la *localisation* de données dans un fonds (fermé ou ouvert) ;
2. la *description*, la *classification*, l'*analyse*, l'*interprétation*, … de données localisées ;
3. l'*éditorialisation* (publication/republication, visualisation, …) et l'utilisation de données préalablement localisées et analysées.

Si nous prenons l'exemple des fonds de données composant les archives audiovisuelles, un des problèmes les plus ardus est celui des *modèles* qu'il faut utiliser pour interpréter – classer, indexer, commenter, comparer, … – des corpus choisis de ces données afin d'expliciter leur *valeur* (i.e. leur sens) pour un acteur social donné. Ces modèles doivent à la fois tenir compte des *contraintes structurales* des données textuelles et des usages pour lesquels ces données sont prévues. C'est ici où la sémiotique – et je pense plus particulièrement à la *sémiotique structurale* – peut jouer un rôle pivot.

La question des modèles de description des corpus de données prend place dans le cadre des recherches théoriques et appliquées consacrées aux *ontologies* au sens informatique du terme, i.e. aux *structures conceptuelles* qui sous-tendent un modèle ou une bibliothèque de modèles de description. Les ontologies sont des *vocabulaires*, des *terminologies* qui expriment ou désignent des *notions*, des *concepts* relatifs à un *domaine de connaissances* et s'adaptent à la *culture (l'expertise) de leurs auteurs* et aux contextes d'usage auxquels ils sont destinés. Les activités de conception, développement, validation et enrichissement d'ontologies font partie du domaine de l'*analyse et de la représentation de connaissances* et s'apparentent à maints égards à celles de l'analyste de corpus de textes *lato sensu* et aussi à celles du lexicologue-terminologue.

Aujourd'hui, on peut constater une tendance à la prolifération incontrôlée de telles structures conceptuelles, ce qui pose toute une série de questions scientifiques et techniques extrêmement épineuses parmi lesquelles on peut compter :

- la portée et la valeur empirique d'une ontologie ;
- les principes de construction d'une ontologie ;
- la structure canonique qui « fonde » une ontologie ;
- le degré de généricité d'une ontologie ;
- la « portabilité » d'une ontologie (i.e. la possibilité de l'utiliser en dehors du domaine pour lequel elle a été conçue) ;
- l'identité et le statut des *catégories conceptuelles de base* d'une ontologie, voire de toutes les ontologies ;
- la distance conceptuelle entre deux ou plusieurs ontologies et la traductibilité d'une ontologie à une autre ontologie ;
- l'intégration, dans une ontologie spécialisée, de parties d'un langage « populaire » (exemple : l'intégration de certaines parties pertinentes de RAMEAU de la BnF[4] dans une ontologie spécialisée sur la diversité ;
- la prise en compte de standards dans une ontologie (par exemple, la prise en compte du formalisme OWL[5] dans l'élaboration d'une ontologie spécifique, la prise en compte de standards communs tels que DCMI[6] ou EAD[7] etc.) ;
- la motivation – linguistique, cognitive, pragmatique, formelle, … – d'une ontologie (i.e. l'organisation d'une ontologie est-elle « simplement » justifiée du point de vue de son utilisation, reflète-t-elle des visions culturelles ou scientifiques particulières ; est-elle motivée par une catégorisation lexico-sémantique proche d'une langue naturelle, … ?) ;
- la dimension de l'évolution des concepts d'une ontologie.

---

[4] *cf.* à ce propos le site https://www.bnf.fr/fr/bnf-datalab

[5] Cf. Lalande, Steffen ; Beloued, Abdelkrim et Stockinger, Peter : Modélisation et formalisation RDF-S/OWL d'une ontologie de description audiovisuelle ; in/ Les Cahiers du Numérique 11/3, 2015, p. 39 – 70

[6] Dublin Core Metadata Innovation : http://dublincore.org/

[7] Encoded Archival Description : https://www.bnf.fr/fr/ead-encoded-archival-description

Enfin, la question de la valorisation, de la création de chaînes de valeurs autour d'un ensemble de données potentiellement signifiantes pour un acteur social, est étroitement liée à celle des *données ouvertes* (« open data »). Celle-ci renvoie à différents problèmes parmi lesquels on trouve d'une manière récurrente :

– le problème de l'accessibilité des données (numériques) au sens *technique* du terme ;
– le problème de la conservation à long terme des données d'intérêt général (par exemple, des données scientifiques, des données du patrimoine culturel, …) ;
– le problème de l'enrichissement des données, de leur utilisation ou réutilisation dans des contextes et cadres d'activités spécifiques.

## 5) L'INSTRUMENTATION DU TRAVAIL AVEC UN CORPUS DE DONNEES TEXTUELLES

Il s'agit des environnements et outils – notamment informatiques – indispensables pour produire, gérer, exploiter et, enfin, archiver et transmettre des données, des fonds de données et, enfin, des corpus de données à proprement parler. C'est un domaine d'expertise à part entière qu'il devient de plus en plus difficile de comprendre, sans parler de sa maîtrise technique à proprement parler. Pour les SHS en France, c'est, bien sûr, le CNRS qui joue un rôle central dans la mise en place de cet écosystème numérique indispensable à tout projet de recherche. Citons à titre d'exemple, les deux TGIR Humanum et Progedo qui s'efforcent d'offrir toute une série de services de base à la communauté scientifique. Nous pensons aussi, bien évidemment, à l'archive ouverte HAL qui est devenu un acteur tout à fait central pour la recherche.

Dans le cadre de cette journée d'études plusieurs communications ont fait référence ou s'appuient directement sur un environnement de travail conçu et développé par les deux chercheurs de l'Ina (Institut National de l'Audiovisuel), Steffen Lalande et Abdelkrim Beloued. Okapi (= « *Open Knowledge Annotation and Publishing Interface* ») offre un environnement très sophistiqué qui permet aussi bien la constitution, l'analyse et la publication de corpus de données multimédias que la modélisation sémantique de domaines de connaissance sous forme d'ontologies de domaine, de thésaurus et de modèles d'analyse et publication (Beloued et Lalande 2017).

## EN GUISE DE CONCLUSION…

Pour terminer, il nous semble que ces quelques thématiques transversales aux contributions réunies dans cet ouvrage mettent en avant, in fine, les trois questions centrales suivantes

1. Comment expliciter la ou les valeurs, i.e. le *sens* de ces données (problème auquel s'intéresse surtout les recherches théoriques et appliquées sur les *« smart data »* et les *« linked data »*, i.e. le *Web des données*) ?

2. Comment modéliser et instrumenter les stratégies d'appropriation/de réappropriation, de publication/republication et, enfin, d'utilisation/réutilisation de ces données par des *acteurs* et dans des *contextes sociaux* les plus divers) ?
3. Comment modéliser et instrumenter les cycles de production, gestion, analyse, diffusion, réutilisation et conservation des données (= question des nouveaux écosystèmes sémiotiques et culturels de communication *lato sensu*).

Ces trois questions renvoient toutes à la *même problématique centrale* qui est celle de savoir « quoi faire avec (ces données) », i.e. comment *appréhender leur sens*, comment les *rendre signifiantes*, comment en faire – pour parler avec le sémioticien A.J. Greimas – des *objets de valeur*.